\documentclass[manuscript]{acmart}
\AtBeginDocument{%
  \providecommand\BibTeX{{%
    \normalfont B\kern-0.5em{\scshape i\kern-0.25em b}\kern-0.8em\TeX}}}
  \usepackage{csquotes}
  \usepackage{graphics}
  \usepackage{csquotes}
  \usepackage[normalem]{ulem}
  \usepackage{xcolor}

\setcopyright{acmcopyright}
\copyrightyear{2018}
\acmYear{2018}
\acmDOI{10.1145/1122445.1122456}

\acmConference[CSCW '23]{Proc. ACM Hum.-Comput. Interact. CSCW 23}{October 13--18, 2023}{Minneapolis, MN}
\acmBooktitle{CSCW '23: Proc. ACM Hum.-Comput. Interact.,
  October 13--18, 2023, Minneapolis, MN}
\acmPrice{15.00}
\acmISBN{978-1-4503-XXXX-X/18/06}

\begin{document}

%\title{"Hey, Can You Add Captions?": Infrastructuring for Community Creative Support on TikTok}
%\title{"Hey, Can You Add Captions?": Critical Infrastructuring for Accessibility on TikTok}
\title{"Hey, Can You Add Captions?": The Critical Infrastructuring Practices of Neurodiverse People on TikTok}

\author{Ellen Simpson}
\email{ellen.simpson@colorado.edu}
\orcid{0000-0003-0387-7329}
\affiliation{%
 \institution{University of Colorado Boulder, USA}
 \streetaddress{1045 18th Street, Campus Box 315}
 \city{Boulder}
 \state{Colorado}
 \country{USA}
 \postcode{80309-0315}
 }

\author{Samantha Dalal}
\email{samantha.dalal@colorado.edu}
\orcid{}
\affiliation{%
 \institution{University of Colorado Boulder, USA}
 \streetaddress{1045 18th Street, Campus Box 315}
 \city{Boulder}
 \state{Colorado}
  \country{USA}
 \postcode{80309-0315}
 }

\author{Bryan Semaan}
\email{bryan.semaan@colorado.edu}
\orcid{0000-0003-1151-2389}
\affiliation{%
 \institution{University of Colorado Boulder, USA}
 \streetaddress{1045 18th Street, Campus Box 315}
 \city{Boulder}
 \state{Colorado}
  \country{USA}
 \postcode{80309-0315}
 }

\renewcommand{\shortauthors}{Simpson et al.}

\begin{abstract}
Accessibility efforts, how we can make the world usable and useful to as many people as possible, have explicitly focused on how we can support and allow for the autonomy and independence of people with disabilities, neurotypes, chronic conditions, and older adults. Despite these efforts, not all technology is designed or implemented to support everyone's needs. One technology that has recently been in the press for how challenging it can be for some creators to use is the short-form video sharing platform, TikTok. Recently, a community-organized push by creators and general users of TikTok urged the platform to add accessibility features, such as closed captioning to user-generated content, allowing more people to use the platform with greater ease. Our work focuses on an understudied population--people with ADHD and those who experience similar challenges--exploring the creative practices people from this community engage in, focusing on the kinds of accessibility they create through their creative work. Through an interview study exploring the experiences of creatives on TikTok, we find that creatives engage in critical infrastructuring--a process of bottom-up (re)design--to make the platform more accessible despite the challenges the platform presents to them as creators. We present these, bottom up, critical infrastructuring practices through the themes of: creating and augmenting video editing infrastructures and creating and augmenting video captioning infrastructures. We then reflect on how the introduction of a top-down infrastructure - the implementation of an auto-captioning feature - shifts the critical infrastructure practices of content creators. Through their infrastructuring, creatives were revising the broader sociotechnical capabilities of TikTok to support their own needs as well as the broader needs of the TikTok community. We discuss how the routine of infrastructuring accessibility is actually best conceptualized as incidental care work. We further highlight how accessibility is an evolving sociotechnical construct, and forward the concept of contextual accessibility.
\end{abstract}



\maketitle

\section{Introduction}
Accessibility describes how we can make the world as usable and useful by as many people as possible \cite{iwarsson2003accessibility,story2001principles}. Accessibility efforts are everywhere, we see them in how people work to ensure people can access and use everything from public transit to online webpages and digital platforms \cite{Morris2016, Qadari2020}. At the heart of many conversations about accessibility is a normative idea that improving access to people, places, and spaces through technology will enhance the quality of life for people with disabilities, neurodiverse people, older adults, and people with chronic health conditions, by allowing for independence and autonomy \cite{InterdependenceAsstTech, AccessibilityBaseddesign, rosqvist2020neurodiversity}. We draw on this broad definition to highlight the breadth of experiences people have around their ability and the range of technologies designed to support people in independently enacting their everyday routines. This is not a paper about assistive technology---any technology that enhances or maintains the functional capabilities  of people with disabilities, neurodiverse people, older adults, or people with chronic disabilities \cite{ITAccessibilityWorkforce, rosqvist2020neurodiversity}--- rather, this is a paper about everyday experiences working with a particular technology. 

We focus on the experiences of neurodiverse people interacting with technology to do creative work, as well as build and find community around that work. Neurodiverse describes the diverse neurological makeups of people across populations shapes how people think and process and how they routinely interact with the world \cite{rosqvist2020neurodiversity}. Neurodiverse often refers to people with autism, dyslexia, or Attention Deficit Hyperactivity Disorder (ADHD) \cite{rosqvist2020neurodiversity}.

This paper explores the experiences of neurodiverse people, but focuses specifically on people with ADHD, as well as the people whose content creation processes tacitly support them. In HCI, research in collaboration with the neurodiverse community has largely focused on people - in particular children - with autism \cite{Spiel2019autistickids, SpielHCIGamesND}, and considerably less inquiry has taken place in collaboration with the adult ADHD community that focuses specifically on their everyday experiences with technology while also having ADHD \cite{SpielADHDTech2022}. This is of note because according to the National Institute of Mental Health, there is an estimated 4.4\% of the U.S. adult population have some form of ADHD \cite{kessler2006prevalence}. Given how prevalent ADHD is, the lack of research attention paid to those with ADHD is concerning. Prior HCI work on ADHD tends to adopt a medicalized standpoint \cite{SonneADHDAssistive} leading to top-down interventions that frame ADHD as a deficit \cite{SpielADHDTech2022}.

The invisibility of adult ADHD leads to a lack of infrastructural support for people with ADHD (or other neurodivergences) across many online platforms, and the mobilization of individuals to adapt and repurpose infrastructures to work for them \cite{star1996steps}. These \textit{critical infrastructuring} practices are the bottom-up actions of communities as they develop supplemental support systems by bridging, filling, or stitching together disparate infrastructures, allowing them to prioritize and imbue the infrastructure with values that reflect the needs, norms, and values of the community as a whole \cite{Britton2019}. These practices push back against the top-down norms and prioritization of the infrastructures that scaffold our everyday routines \cite{star1996ecology}, such as how polio survivors modify or adapt assistive technologies like replacement limbs or braces to support their needs \cite{hamraie2017building}. Studies examining critical infrastructuring have looked at the needs of postpartum mothers \cite{Britton2019} and polio survivors \cite{hamraie2017building}, less scholarly attention focuses on the practices neurodiverse people, particularly those with ADHD, to make infrastructures work for them. In this paper, we shift attention away from this deficit framing to explore the everyday infrastructural experiences of neurodiverse people. \par

We explore these infrastructural experiences by focusing on a platform that is increasingly growing in popularity--TikTok. Recently, TikTok has come under scrutiny for it's treatment of people of different access needs as TikTok breaks many commonly understood design principles around ease of use and universal design \cite{Kobie2019}, rendering some parts of the platform inaccessible to some users \cite{Odell2020}. In recent years, there has been a community-based push by video makers and viewers across social media to add accessibility features such as closed captioning to user-generated content because audio-based formats are simply not accessible to many users, such as those who are d/Deaf or hard of hearing, or those who have attention or auditory processing issues such as those with ADHD or Autism \cite{Lerman2021, Odell2020}. On TikTok, this user-driven effort was successful, resulting in TikTok implementing an auto-captioning feature in April 2021\footnote{https://newsroom.tiktok.com/en-us/introducing-auto-captions}. This community-driven push for accessibility on TikTok is especially important when considering how TikTok platform has become an important site for information access and grassroots activism around social, environmental, and political issues \cite{compte2021s, Serrano2020, Allaire2020}. For the purposes of this paper, we explore how neurodiverse creatives on TikTok, particularly those who report having ADHD, routinely engage in critical infrastructuring practices to extend the ease of use and access both their creative work and TikTok as a whole. \par

Drawing on 15 semi-structured interviews with TikTok video makers, as well as five follow-up conversations, this paper examines the critical infrastructuring practices of people who create and share "accessible" videos on and for TikTok. This paper explores accessibility in the community-led, bottom-up sense, where accessibility is about creating access and is derived from community norms that are negotiated through social interactions \cite{robinson2018phenomenological, Low2019twitter, 2016ComicTranscriptionsFiesler}.  We find that creatives on TikTok integrated critical infrastructuring practices into their routine creative work to address accessibility issues and creating access on the platform. Participants described: creating and augmenting video editing infrastructures, and creating and augmenting captioning infrastructures. They then reflected on the tensions that emerge when a top-down infrastructural solution, auto-captioning, is implemented. Participants discussed their loss of ability to build routine care into the videos they created, such as ensuring accuracy and clarity, while also the loss of the ability to design their videos in a way that made their captioning work legible to users. Through their critical infrastructuring, creatives were revising the broader sociotechnical capabilities of TikTok to both create accessibility around ADHD and other neurodivergence that supported their own creative efforts while also creating accessibility that supported the technological experiences of the broader user-base. We discuss how, in this case, the routine of critical infrastructuring is best conceptualized as accidental care work. We highlight how accessibility is an evolving sociotechnical construct, and foreground the concept of contextual accessibility. We conclude by discussing implications for design and future work that can continue to inform design practices. \par

\section{Related Work}
We discuss existing literature around routines, and note the relationship between routines and creativity. We turn to a discussion of the role infrastructure plays in enacting everyday routines. When infrastructure is incomplete, or breaks down, people's everyday routines are disrupted, leading them to draw on other infrastructures in order to repair or complete these infrastructures. We briefly touch technology and neurodiverse people, the creative work of neurodivergent content creators on social media, before turning our attention to the case of User-Generated Content (UGC) and the critical infrastructuring practices required to create accessible UGC in incomplete platform infrastructures. \par

\subsection{Routines and the Infrastructures that Support Them}
At the core of many questions about technology use are routines. As people go about their lives, they develop routines, recognizable patterns of behavior or action that are carried out by multiple actors within a specific context \cite{feldman2000organizational}. Recent studies reveal that routines can be adaptable and can function outside of behavioral rules and norms \cite{pentland2012dynamics}. The dynamics of these routines show the agency people have in both forming new routines and modifying existing ones. \par

Routines are independent patterns of actions \cite{pentland2012dynamics}. These systems of actions have two interacting parts: the ostensive and the performative\cite{pentland2012dynamics}. Here, the ostensive part of a routine is the higher level understanding of a specific pattern of action which serves as a guide for a specific performance of a particular routine. The performative aspect of a routine, conversely, is the contextual performance of "specific people at specific times, in specific places" \cite{Semaan2019}.  For example, a person can perform the ostensible routine of preparing to create art in multiple different ways. They could gather their materials and prepare their piece by applying gesso to a canvas, selecting a piece of wood to turn on a lathe, or stenciling an outline onto metal to cut out. Conversely, another person may spend time on Google Image Search looking for digital references to use while drawing in an art program like Procreate. In essence, how a person performs any given routine pattern of behavior can be different, depending on its situational context.\par

In order to enact their everyday routines, people often draw on infrastructure. Infrastructures are the underlying foundation of any large-scale system society replies on to support routine functions \cite{edwards2003infrastructure}. For example, city roads are an infrastructure that allow people to routinely travel to work daily. While infrastructure is largely understood as both physical and technical, information and communication technologies are also infrastructure \cite{hanseth2010design}. \textit{Information infrastructure} comprises information systems like databases and internet-enabled systems like social media platforms, which serve as the installed-base of IT capacities to deliver and allow a community to use information services \cite{hanseth2010design}. Information infrastructures also allow individuals, groups, and societies to enact everyday routines, as they are integral to routine communication and collaboration between individuals. \par

Infrastructures, both physical and digital, are sociotechnical systems \cite{edwards2003infrastructure}. They shape and are shaped by social practices \cite{star1996steps}. Physical and technical infrastructures are often intertwined with human social organization and practices \cite{star1996steps}. Therefore, infrastructure is defined in use--it takes on meaning or changes in meaning depending on the social practice taking place and the actors involved. Put another way, infrastructure is a continually negotiated relationship between the people and contexts involved; they are relational systems \cite{star1996steps}. \par

Infrastructures comprise both technological components and physical components but are given meaning by the social practices in which they are enacted. \textit{Human infrastructure,} “the arrangements of organizations and actors that must be brought into alignment in order for work to be accomplished.”\cite{LeeDourishMark2006}. Human infrastructure allows the work of our everyday routines to take place; they animate and sustain the infrastructures that enable societies to function. Human infrastructure can be a combination of entities that are both known and unknown \cite{LeeDourishMark2006}. \par 

\subsubsection{Creative Routines}We are interested in how creatives draw on digital infrastructure to engage in the ostensible routines around creativity. The routine, everyday practices of creativity to produce cultural artifacts like tweets, pieces of fan art, or TikTok dances, are a form of vernacular creativity, where the act of cultural production is in the hands of ordinary people \cite{burgess2007vernacular}. Online, cultural producers, individuals who create, share, market, and monetize creative works, engage in routine, creative practices enacted on and drawing upon, platform infrastructures \cite{poell2021platforms}. In HCI, creative routines and creativity itself are bounded up in the ability to \textit{make things} \cite{dogross2007creativity}. For example, people with physical disabilities after surviving polio engaged in routine creative practices to create infrastructures of and tools for accessibility for themselves when those infrastructures did not yet exist in their everyday lives \cite{hamraie2017building}. Creative routines are also important in articulating a particular identities. Fathers who run do-it-yourself blogs engage in creative practices allowing them to articulate identities through a confluence of material and cultural co-production \cite{Ammari2017Fathers}. The ordinariness of these creative routines all rely on there being an existing infrastructure upon which to draw upon. If the existing infrastructure is incomplete or simply not there, or has somehow broken down, engaging in everyday creative routines can be challenging. Next, we consider how such breakdowns or incomplete infrastructures can impact creative routines.

\subsection{Infrastructural Breakdowns, Incomplete Infrastructures, and the Ways People Mitigate Them}
The routine functions of digital infrastructures rely on the emergence of new routines around creative labor. Infrastructures involve many interconnected parts, but they are often also invisible---operating in the background and allowing people to engage in their everyday routines without paying them much mind \cite{star1996steps}. They only become noticeable when they fail to meet peoples needs in the form of an \textit{infrastructural breakdown} \cite{star1996steps}, or when they routinely disrupt people's lives \cite{Semaan2019}. The disruption of infrastructural breakdowns is not always universal \cite{star1996steps}. Infrastructures are not value-neutral, as it is the invisible human actors--the designers, curators, and maintainers--that make these infrastructures work \cite{bowker1994information}. The values embedded into infrastructures can be the root of routine sources of disruption in people's everyday lives \cite{Semaan2019}, such as when a video-sharing platform like TikTok or YouTube does not offer automated closed-captioning or subtitling services. The platform infrastructure supports people who can hear and process audio, while not supporting the needs of people who have auditory processing or attention issues, or are neurodiverse. In some cases, value misalignment between designers and users can result in \textit{incomplete infrastructure}---an infrastructure that does not meet the needs of those who depend upon it to enact their routines. \par

Breakdowns surface both underlying and taken-for-granted infrastructures and the invisible work needed to maintain and sustain infrastructures \cite{star1999ethnography, bowker2000sorting}. When faced with incomplete infrastructures or to mitigate infrastructural breakdowns, people engage in \textit{infrastructuring}---the intentional bottom-up building of infrastructure to generate the supplemental support systems required to meet their needs\cite{star1996steps}. Infrastructuring can fulfill many interrelated goals. This process of bridging, or filling, an infrastructural gap \cite{EriksonJarrahi2016, erickson2014more} can involve stitching together multiple disparate infrastructures, drawing attention to what is and is not supported by existing infrastructures through a process of stitching together "fleeting, nonstable, even ephemeral moments of alignment" \cite[~p.277]{vertesi2014seamful, bowker2000sorting}. This practice allows for us to view the collective work of actors to produce moments of seamlessness, where the infrastructure once again serves its desired purpose \cite{vertesi2014seamful}. For example, on Reddit, community moderators mitigate Reddit's lack of infrastructural support for maintaining safe communities by drawing on external resources, such as auto-moderation bots, to lessen the impact of the emotional labor that enforcing community rules and norms has on them \cite{dosono2019moderation}. The curation and development of a suite of external tools demonstrates how these moderators engaged in invisible labor to make seamless their everyday routine of content moderation more accessible and less emotionally taxing \cite{dosono2019moderation}. 

The practice of stitching disparate technologies together to create a seamless, functioning infrastructure demonstrates, too, how individuals push back against the values embedded within infrastructures. Infrastructures by nature are top-down, so the values of the designers are embedded into infrastructures in ways that inevitably lead to tension when designers do not share the values as members of the user base for whom they have built infrastructure. Infrastructures are an inherently political technology \cite{winner2017artifacts}, and considering infrastructure from a bottom-up perspective focuses on how communities of disparate individuals find, engage, and build with each other, oftentimes through the medium of online platforms \cite{Britton2019, gray2019ghost, Ammari2017Fathers}. Findings from such studies demonstrate to how infrastructure is imbued with the values of those who co-construct it, therefore the \textit{critical infrastructuring practices}---the bottom-up practices of communities as they develop supplemental support systems by bridging, filling, or stitching together disparate infrastructures---of communities allow them to imbue the infrastructure with values that reflect community norms and values \cite{Britton2019}. This makes bottom-up infrastructures potentially more resilient, flexible, and appropriate for communities in which norms and values are not static \cite{Semaan2019}. With critical infrastructuring in mind, we turn our attention to the relationship between neurodivergent people and technology. \par

\subsection{Neurodivergence, ADHD, and Technology}
Neurodiversity refers to how multiple neurological makeups of people across populations shapes and impacts the way they think and process information, as well as their routine engagements with the the world  \cite{rosqvist2020neurodiversity}. The term neurodiversity emerged from self-advocacy from people with diverse neurotypes, such as people with autism, ADHD, or dyslexia, who rejected the labeling of specific cognitive processing styles as an impairment or deficient in some way \cite{SpielADHDTech2022, rosqvist2020neurodiversity}. While these terms are medical labels, neurodiverse people have many different relationships to and with their experiences with neurodivergence \cite{SpielHCIGamesND} and sometimes medical labels cast people with specific neurotypes as "disabled or devalued" within neurotypical society \cite{chapman2022neurodiversity}. While neurodiversity typically used to describe this multitude of experiences, \textit{neurodivergence} is the "experience of significant difference from what is understood as the norm of cognitive functioning and expression (neurotypical)" \cite[p.~1]{SpielADHDTech2022} \cite{rosqvist2020neurodiversity}. \par

Calls to work collaboratively with neurodivergent people in HCI to develop assistive technologies that serve the community's needs are an opportunity for HCI as accessibility research and design are "dominated by neurotypical thinking" \cite[p.~2302]{Dalton2013NDandHCI}. Much of the HCI research on accessibility for neurodivergent people focuses on people with autism \cite{SpielHCIGamesND, williams2020perseverations}, particularly around autistic children \cite{Spiel2019autistickids}. A 2022 literature review by a team of neurodivergent authors \cite{SpielADHDTech2022} notes that many of the assistive technologies introduced to HCI for neurodivergent people are largely used to mitigate the subject's experiences of neurdivergency, which is considered to be disruptive to neurotypical people's everyday experiences. \par

While we discuss creatives with different divergences in this paper, as well as those whose creative work supports the accessibility needs neurodivergent people, we primarily focus on people with ADHD. Medically, ADHD is characterized by traits such as hyperactivity, impulsivity, and inattention \cite{epstein2013changes}. ADHD is typically diagnosed during childhood \cite{singh2015overview}, and often seen through a gendered \cite{horton2018voices} or racialized \cite{SpielADHDTech2022,ballentine2019understanding} lens. In popular culture, and in some medical circles, ADHD is a childhood disorder that is only common in boys that they grow out of \cite{shah2013adults, asherson2012adultadhd, horton2018voices}. \par

According to the National Institute of Mental Health, about 4.4\% of American adults are estimated to have a current diagnosis of ADHD \cite{kessler2006prevalence}. Adults with ADHD are more likely to experience disruptions in their everyday routines because of their ADHD. Research in psychology shows that adults with ADHD are often under-employed or unemployed at much higher rates than their neurotypical counterparts due to struggles with sustaining attention and following through with assigned tasks disrupting their ability to routinely complete tasks required for work \cite{murphy1996attention}. People with ADHD also underachieve academically in post-secondary education due to problems with task prioritization and lack of accommodation from universities, which routinely disrupts their ability to learn and achieve scholastically \cite{biederman2008educational, advokat2011college}. Other psychology studies show how ADHD is disruptive for people's interpersonal \cite{murphy1996attention, biederman1993patterns}, romantic, and family relationships \cite{harpin2005effect, eakin2004marital}. Historically, psychology research did not qualitatively draw on the individual experiences of adults with ADHD, however there is a growing body of research that seeks to draw on these lived experiences to better understand the everyday experiences of adults living with ADHD \cite{horton2018voices,schrevel2016need}. The new push to prioritize the experiences of adult individuals who have ADHD demonstrates the invisibility of ADHD in adults, as well as the lack of research into how ADHD can disrupt the everyday routines that prioritizes the voices of people with ADHD. It further emphasizes  as well as the lack of recognition and accommodation of ADHD as a disability despite being identified by the Americans with Disabilities Act\cite{asherson2012adultadhd,heekin2010adhd}.  \par

As information about ADHD becomes more available to people online, efforts to communicate the everyday experiences of having ADHD as creative people by people with ADHD have become popular. Widely shared pieces of art, such the work of artists like Pina Varnel\footnote{See a list of Varnel’s work here: https://twitter.com/ADHD\_Alien} or Dani Donovan on Twitter\footnote{See Donovan’s website here: https://www.adhddd.com}, or videos like those created by Jessica McCabe on YouTube\footnote{See McCabe’s YouTube channel here: https://www.youtube.com/c/HowtoADHD} provide just such a window, with helpful anecdotes and strategies for coping with having ADHD -- particularly as an adult who engages in creative work. These cultural artifacts serve as a pop culture-driven answer to the overall lack of research that centers and prioritizes the voices and experiences of people with ADHD \cite{horton2018voices,SpielADHDTech2022,schrevel2016need}. These technology-mediated cultural artifacts speaking to the everyday experiences of having ADHD however, are not always treated seriously by the research community. A recent review of videos shared on TikTok about ADHD identified that of the top 100 posts tagged \#ADHD, only 11 were shared by credentialed health care providers, and 52 of the videos were classified as misleading in some way \cite{yeung2022tiktok}. A review of YouTube produced similar results \cite{thapa2018youtube}. While these videos were deemed "misleading," many of them were classified as such due to the use of non-medically specific terms (e.g. "time blindness" as opposed to the medically-recognized term, "inattentiveness"). Both internal to and external to the HCI community, a great deal of the research available tends to frame ADHD through a medical--as both Yeung and colleagues' \cite{yeung2022tiktok} and Thapa and colleagues' \cite{thapa2018youtube} analyses do--or through a deficit lens \cite{SpielADHDTech2022}. The use of non-medically specific terms demonstrates how people with ADHD are creating and sharing information about their everyday experiences with ADHD are working to make their content accessible---easily digested and understood---by the people who consume the cultural artifacts they produce. This practice also forwards a secondary question: How do people with ADHD experience their neurodivergence during the course of their everyday lives and routines around being creative, either about their ADHD or just in general? In the next section, we review the practice of online content creation and discuss our research questions.\par

\subsection{Critical Infrastructuring to Support Content Creation for Neurodiverse People}

Online social platforms rely on user-generated content (UGC) to sustain themselves \cite{bruns2009prosumer}. Examples of these platforms include YouTube, Twitter, TikTok, Twitch, and countless others. By considering the production of UGC as labor, a platform's userbase become “prodsumers” of the platform, as they participate in the consumption of the products they co-create \cite{toffler1980third}. These platforms also depend on their userbase to not only create, share, and consume UGC, but also collaboratively generate the knowledge needed to sustain the platform with other users. This renders users not only prodsumers of a platform, but also a \textit{"produsers"} \cite{bruns2006towards}. Produsers are essential to a platform's functionality, as they, through their labor, make the platform complete. Without users, a platform has no value, as there is no content to share. Individual users become a part of the platform's human infrastructure, animating otherwise incomplete platforms through routine creative labor. \par

The essential role of produsers in maintaining platforms is well-documented. On Instagram \cite{van2021selling}, TikTok \cite{klug2020took}, and YouTube \cite{bishop2020algorithmic}, previous work has shown how reliance on UGC introduces an element of professionalization to produsers who do creative labor on these platforms, while also acknowledging the precarity of this creative work due to platform governance and infrastructural practices \cite{duffy2021nested, poell2021platforms}. Through their creative labor, produsers engage in infrastructuring work, drawing on the knowledge of the collective to better understand the logics of the platform where they create, produce, share, and consume cultural artifacts \cite{bishop2019managing}. These human knowledge infrastructures allow produsers to patch the perceived fissures in platform infrastructures following moments of breakdown \cite{Semaan2019, Qadari2020}. \par

There is a wealth of research showing how produsers of particular spaces, places, or platforms engage in infrastructuring practices to address the infrastructural breakdowns they experience.  This paper builds on this prior work by examining how produsers engage in infrastructuring to address perceived infrastructural gaps within the context of both their everyday use of, as well as their work creating and sharing UGC on TikTok from both a social and technical perspective. We are particularly interested in the work that neurodivergent TikTok content creators and those who tacitly support them do to make their content more accessible to people of all neurotypes, especially to those with ADHD. Previous work explores the development of crowd-sourced solutions for inaccessible UGC, such as community-driven efforts to create transcriptions of comic books within a fan community \cite{2016ComicTranscriptionsFiesler}. Other work examines how online platforms do not support creators in allowing their creative work to become visible \cite{bishop2019managing} or by making their volunteer work challenging through sweeping decisions regarding platform infrastructures \cite{dosono2019moderation}. To date, however, little work directly examines how produsers themselves experience the lack of infrastructural support from the platforms where they enact their creative routines, and how they alter their routine practices around producing and sharing in creative work to make their creative work more accessible---consumable for people with different abilities and neurotypes.\par

This paper examines bottom-up norm construction on the video-sharing platform, TikTok, exploring the tensions between TikTok's top-down infrastructure---as a platform designed and maintained by a private company with its own incentives for it's infrastructural design choices---and the diverse and diffuse userbase of produsers who define their social norms in a bottom-up, community-led manner. The tension between a top-down infrastructure design and bottom-up social norm construction, particularly around issues of accessibility for neurodivergent people, surfaces an infrastructural breakdown of TikTok. The incongruities between social norms around accessibility and the prevalence of inaccessible videos has driven recent conversation concerning the lack of accessibility of TikTok \cite{Odell2020}.\par 

Our work focuses directly on the creative practices of neurodiverse people, and how they engage in critical infrastructuring work as a part of their everyday creative routines on TikTok to address these perceived infrastructural gaps. Examining TikTok's infrastructure, our study participants found aspects of it inaccessible, therefore disrupting their established creative routines. In order to fill these infrastructural gaps to continue to enact their creative routines, video creators on TikTok engage in additional \textit{creative labor}, defined here as the time, effort, and creativity needed to produce an artistic output. When TikTok's infrastructure fails to provide the necessary tools to render its hardware, platform and community accessible to all it's users, videos, the produsers engage in critical infrastructuring to repair this infrastructure and meet these commitments. To date, as Spiel and colleagues \cite{SpielADHDTech2022} point out, there are very few papers within HCI that directly reflect on the everyday experiences of people with ADHD using technology. We seek to contribute to this space by discussing how content creators with ADHD, as well as those who share similar challenges or wish to support them, engage in critical infrastructuring practices to draw on disparate infrastructures to create a seamless experience of both content creation and content consumption by others. Thus, our work moves from a deficit model to one exploring the bottom-up, intentional, production of infrastructure amongst people neurodiverse people. \par

\section{Research Methods}
These findings come from a larger study examining the experiences and practices of people who create and share TikTok videos. This project was originally focused on understanding people's creative routines on a platform like TikTok, we did not realize we would be working with a population of creatives that was largely neurodivergent (n=8/15 participants specifically stating their neurodivergence). Yet, during our discussions, concerns around accessibility in both the technical and social sense, as well as our participant's neurodivergence,or experiences they shared with neurodivergent participants featured heavily. Participants discussed how they looked outside of TikTok to address these concerns around their neurodivergencies and those of their friends and audiences, altering their creative routines to address their accessibility needs which led to them also addressing the accessibility concerns of others. This organizing theme around critical infrastrucuturing was identified through an inductive analysis of the interview text. Participants described drawing on multiple infrastructures to make TikTok \textit{usable} for both themselves and people like them. These interviews were semi-structured, and designed to serve as life histories \cite{wengraf2001qualitative}. They were conducted between February and March 2021, after being approved by our institution's Institutional Review Board. 

\subsection{Participants and Recruitment}
Recruiting individual study respondents directly from TikTok is challenging, but direct solicitation of individuals in the form of recruitment videos has had success \cite{foryouforyou}. The first author made a recruitment video and shared it on their personal TikTok introducing the research and it's general goals, following Simpson and Semaan's method \cite{foryouforyou}. The video invited participants to complete a recruitment survey shared in the comments. A call for respondents was shared on the first and third author’s social media accounts. Further recruitment took place in two private discord communities for TikTok Creators with permission from the community moderation teams. Following each interview, we used snowball sampling to recruit further participants \cite{biernacki1981snowball}. By triangulating across multiple recruitment sites, we hoped to avoid sampling bias. This approach has been used by other HCI studies (e.g. \cite{hagar2005crisis}). Interested participants were provided with a link to a Qualtrics recruitment survey hosted by our university. This survey collected basic demographic information about participants (e.g. age, gender identity, pronouns, sexuality, and racial identity). Each of these responses was a free-response box, allowing participants to self-identify \cite{JaroszewskiAttackHelicopter}. See Table 1 for a breakdown of participant demographics.

Here, we note that our recruitment documentation did not ask if potential respondents were neurodivergent. Yet, during our interviews, 8 out of our 15 participants disclosed being neurodivergent by specifically naming their neurodivergency (e.g. stating "I have ADHD" [P12], "I'm autistic" [P2]), and 3 others discussed traits that were shared with the neurodivergent participants (e.g. needing subtitles to pay attention to a video [P14], or not being able to concentrate on editing a video long enough to add captions [P6, P15]). Our inductive analysis focused on identifying challenges and traits across our participants, which grouped these 11 participants together at the start of this analysis. Table 1 represents participant demographics as they reported them, with this distinction clearly identified. This inductive finding of an aspect of peoples' identities that we did not actively account for in our recruitment materials is of particular note as the struggles identified spoke to a common community need. \par

Potential participants needed to be (1) over the age of 18; and (2) make videos for TikTok to be eligible for this study. We wanted to capture a breadth of experience in making videos on TikTok that crossed generational lines between Millennials, Gen Z, and Zennials; as previous research \cite{foryouforyou, karizat2021algorithmic} has demonstrated there is a strong presence of these age groups on TikTok. Based on these criteria, the survey responses were subsequently sorted and individual respondents were contacted via email. We reached out to 28 people and received 16 affirmative responses, conducting 15 interviews.

\begin{table}[]
\begin{tabular}{lllll}
P\#  & Age & Gender     & Pronouns  & Neurodivergence                                         \\
1  & 27  & Woman      & she/her   & Not discussed                                            \\
2  & 31  & Woman      & she/her   & Autism     \\               
3  & 30  & Woman      & she/her   & ADHD       \\
4  & 36  & Woman      & She/her   & Neurodivergent                                         \\
5  & 21  & Male       & he/him    & Not discussed                         \\
6  & 29  & Nonbinary  & they/them & Not specifically named \\
7  & 39  & Male       & he/him    & Not discussed                                        \\
8  & 24  & Nonbinary  & they/them & ADHD                                                      \\
9  & 18  & Male       & he/him    & Not discussed                                               \\
10 & 40  & Nonbinary  & she/they  & Autism, ADHD                           \\
11 & 20  & Woman      & she/her   & ADHD                                                \\
12 & 31  & Female     & She/Her   & ADHD                               \\
13 & 31  & Nongender  & Most      & ADHD                                         \\
14 & 31  & Woman      & She/her   & Not specifically named                                  \\
15 & 26  & Non binary & She/They  & Not specifically named                       \\          
\end{tabular}
\end{table}

\subsection{Interviews}
Using a qualitative approach, \cite{strauss1990basics,yin2017case}, we conducted 15 in-depth semi-structured interviews between February 15, 2021 and March 30, 2021. Interviews lasted between 50 minutes and two and a half hours (averaging 100 minutes), with a check-in on the hour to see if the participant wanted to continue. The interviews were conducted over Zoom. Participation was voluntary, and participants were compensated at a rate of \$20.00 USD for up to the first hour, and a subsequent rate of \$20.00 an hour following the first hour. Before starting the interviews, we asked participants if they were in a space where they felt safe to discuss potentially sensitive topics that may come up during the interview, and attained their oral consent to participate in the study. With participant consent, these interviews were audio recorded and transcribed by the first author for analysis.\par 

These interviews were designed as life histories \cite{wengraf2001qualitative}, where participants were asked to guide the interviewer through their early experiences with online community spaces, their creative work on both these platforms and more broadly , on- and off-line. Next, our questions about TikTok were two-fold. We asked general experiential questions about (e.g. how participants found their way onto TikTok, and their general impressions of setting up and  using their For You Page), before moving on to questions specifically about the process of making videos on TikTok. We asked participants to describe their creative process on TikTok, how they stayed engaged with their audience, and where they were struggling with TikTok as creators on TikTok’s platform. In this final section, where participants were asked about their creative process and routines of making videos on TikTok, that the following two questions were asked: "If you could change some aspects of TikTok, what would you change and why?" along with the follow up question of "What about [aspect] on TikTok now is challenging?" From these questions we draw the emergent themes that are at the heart of this paper.\par

Following TikTok's implementation of an auto-captioning feature in April 2021\footnote{https://newsroom.tiktok.com/en-us/introducing-auto-captions}, the research team reached out via email to all 15 participants in June of 2021. In this email we provided participants with copies of their interview transcripts and asked follow up questions to see if participants were using TikTok's new auto-captioning feature, and if not, why not? We additionally asked if TikTok's auto-caption feature satisfactorily provided everything that they would like to see  in a captioning feature, and if not, what would they like to see. Of our 15 participants, five responded (P2, P5, P7, P8, and P10), providing written responses to our questions that are discussed below in Section 4.3.

\subsection{Analysis}
We used an approach based on grounded theory to analyze the interview data \cite{corbin2014basics}, which has been adopted and commonly used by HCI scholars \cite{foryouforyou, karizat2021algorithmic, Britton2019}. The first author used MAXQDA, a qualitative data analysis program, to code the interview transcripts. They conduced a preliminary round of inductive open coding and memoing of the 15 interviews. During this process, the first author met weekly with the third author, discussing emergent themes. Once the open coding was completed, a process of collapsing and refining the emergent themes into categories was conducted by the first author with input from the third author. During this process, responses to the questions listed in the previous section around challenging aspects of TikTok and changes that participants wanted to see on TikTok were largely placed into a broader bucket of 'accessibility concerns,' with other refinement placing coded segments into buckets around visibility, privacy, routine creative processes, and creative burnout. This paper focuses on the first bucket mentioned, as further refinement and sorting showed how participants had established routines around creation and were drawing on multiple different resources both internal to and external to TikTok for both social and technical purposes to create accessibility. Based on these bricolage and stitching tactics, we decided to take an approach rooted in infrastructure theory, considering how participants were creating opportunities by engaging in routine infrastructuring to correct infrastructural gaps they found on TikTok. We address this by examining the tools participants were using to achieve these goals. At this point, the second author joined the project. The first and second author conducted a second round of thematic analysis on the emergent codes around accessibility and the norms around creating videos that are accessible to everyone. We further analyzed the data from the follow-up email responses from P2, P5, P7, P8, and P10 during this analysis pass. From this round of collaborative analysis, the primary themes emerged which we present in the next section. \par

\section{Findings}
We observed three instances where participants found that the infrastructural logics of TikTok did not support their routine creative practices. These failures were rooted in what participants perceived as gaps or incompleteness in TikTok's existing infrastructure that made the platform challenging for them to use. These practices led them to engage in new, ordinary routines around critical infrastructuring. These new routines include video editing infrastructures, infrastructuring care through video captioning, and a reflection on how the loss of control following the implementation of TikTok's top-down infrastructure, auto captioning, presents challenges to these critical infrastructuring practices. Sawyer and colleagues \cite{sawyer2019infrastructural} tell us that information is often fragmented across many different digital platforms, and that infrastructural competence comes from being able to fill infrastructural gaps to establish routine practices seamlessly integrating multiple platforms. In this section we further show how while these bricolage practices allowed our participants to continue to engage in their routine creative practices, they added more labor to these routines. While we found that our participants were creating accessibility in other ways, such as in helping create centralized community spaces that are often disparate, an emergent finding was that the primary focus of their infrastructuring efforts were centered around their experiences and other people's experiences of neurodivergence.\par

\subsection{Infrastructuring for Better Tools: Augmenting TikTok's Video Editing Capabilities with Other Platforms}
In order to cope with the challenge that TikTok's infrastructure presented to their existing creative routines on top of their self-described neurodivergence, our participants discussed how they routinely looked beyond TikTok’s existing app infrastructure in order to find the tools that could help improve their creative routines. In this section we discuss how participants drew upon external infrastructures in order to make videos on TikTok. Oftentimes, our participants cited challenges editing directly within TikTok's internal app infrastructure, which offers a suite of editing tools that allow participants to shoot, clip, and combine recorded video clips. This suite of tools, however, presented challenges related to their neurodivergence, as well as in general, to our participants, as we elaborate in this section. \par 

Our participants came from many different technical skill levels, but all held a general assessment that "better produced" videos on TikTok tended to perform better with audiences. While the creator's intent may be the spark of creativity, only the audience's subjective experience of the creative output is knowable through the platform metrics \cite{poell2021platforms,baym2013data}. Therefore, 'doing well' to our participants is deeply rooted in quantifiable platform metrics. For many, this meant that the videos perceived to be more successful were produced outside of TikTok, rather than within TikTok’s in-app editing suite, which was universally considered by our participants to be hard to use. Some participants (P4, P13) discussed integrating the use of Adobe Premiere, a video editing software they both owned and were familiar with, into their creative routines as they found making videos in TikTok's in-app editing suite challenging on account of their neurodivergency. For example, P4, who discussed her neurodivergence by explaining: "I'm also neurodivergent. So my brain is a little bit nonlinear," describes frustrations engaging with other people's content using TikTok's stitch\footnote{where several seconds of one video are captured before another user’s video is added to the original recording} and duet\footnote{where two videos play side by side in conversation with each other} feature because she was unable to edit them externally to TikTok in Adobe Premiere: \begin{displayquote}P4: "[T]hat is one of the reasons I hate the [...] You know the thing where you, you just grabbed five seconds of someone's video and then you respond to it?" \par
\textit{Interviewer: Stitch.}\par
P4: "Ah, thank you so much. Because stitch you have to do live, which drives me crazy. You can't upload a video to stitch them together, you have to do your response live, you know, just capture it. I never stitch videos, which is really annoying, because I'd love to. [T]here's so many videos, [...] I could stitch up, but I can't because, [...] like I said, because the way my brain works, I can't -- [...] it would be incoherent. If I tried to do like 50 seconds of me just talking; I have to plan it and then I have to edit it down because I usually talk for five minutes and I can only have 30 seconds of content or whatever." \end{displayquote}
When P4 is referring to "live" in the context, she is referring to recording herself in real time, rather than pre-recording and editing that external to TikTok. P4's frustration with her inability to record herself in real time is a common attribute shared by people with many different neurodivergencies \cite{seymour2019frustration, sjowall2013multiple}. This combination of traits makes it difficult to establish creative routines, especially when the infrastructure itself makes it challenging to engage in previously-established routine patterns of creative labor. Coming to TikTok, P4 drew on her existing skills with Adobe Premiere in order to augment TikTok's editing suite, because she struggled to make videos in real time on account of how her neurodivergence led to frustration and P4's inability to record responses that appropriately captured the points she wanted to make. P4's creative routines are incompatible with TikTok's in-app editing software in ways that render some features of TikTok's infrastructure, such as the stitch feature, inaccessible to P4. This inaccessibility makes it so that P4's already augmented creative routines are limited by TikTok's app infrastructure, and certain features and popular creative forms have become closed off to P4 entirely. \par

In order to better understand P4's creative routines on TikTok in general, we asked her to describe her creative routines on TikTok. P4 described the process as follows:
\begin{displayquote}"So I'll capture a self video with my phone. Then I upload that to Dropbox. I'd have to make videos with using the audio download that so that I can import the audio into the video editing software and then re-upload it. I do everything in Adobe. And then I export that back out to my Dropbox. So I can go into Dropbox on my phone and download the video that I made. [...] And then I upload that to TikTok. And then at that point, if I used an audio that I took from TikTok, I go and grab that again from TikTok and I add that [...] so that the right thing is getting credit."\end{displayquote}
P4's creative routine is enacted through bringing together several different technical infrastructures in order to create a single video. She relies on her technical skill using Adobe Premiere, as well as her knowledge of digital storage services like Dropbox, to perform her routine of making a TikTok video that she then shares. These practices also demonstrate why certain aspects of TikTok, such as using the stitch feature, are inaccessible to P4, because she is not able to engage in her creative routines and use these features.\par

The bricolage practices P4 describes help to illustrate how a gap in TikTok's infrastructure, specifically how certain video forms and app features, do not allow for people to engage in their routine creative practices if those routines involve bringing together multiple different technologies. P4's solution, to simply not stitch videos, is one that also highlights how features incompatible with creative routines create access challenges. We see this in how P4 laments that there are "there's so many videos" that she'd "love to" stitch, but cannot because the bricolage practices she's added to her creative routines to accommodate her need to "edit down" what she wants to say on account of her neurodivergence, make certain aspects of TikTok inaccessible. Given that these routines enable P4 to create and share videos on TikTok in the first place, having these routines also render new, different, aspects of the platform inaccessible further demonstrates how accommodation of her needs as a neurodivergent person can be a double edged sword. \par

Editing video on TikTok, in general, was considered challenging by many of our participants. P3, who has professional photo and video editing experience, put it succinctly: 
\begin{displayquote}“If you record video inside of the application, and you want to splice it together in the application, it's just absolute hell, it just does not work.”\end{displayquote}
P3's frustrations with TikTok's editing suite mostly came from her ADHD. Describing her interactions with TikTok's interface and algorithm, P3 explained, "ADHD and this app do not mix well." This, she attributed, mostly to her inability to concentrate on any one task (in this case watching or creating videos) on account of how TikTok allowed her to "channel flip" through various videos -- tying in to how many of the attention issues that neurodivergent people have are directly tied to their short term memory and emerge as "executive dysfunction" \cite{sjowall2013multiple}. P3 identified her attention issues---her inability to track what she was doing from one moment to the next---and attributed them to TikTok's infrastructure, while also describing being unable to concentrate on video editing due to the editing interface's "workflow problems." Here, workflow meant the process of splicing together videos in the editing suite. Because of her professional experience, P3 was easily able to see where the potential problems of editing video directly drawing on TikTok's infrastructure. She voiced a secondary concern as well, indicating that the video quality she was able to upload to TikTok as an Android phone user was not as good as those uploaded from an iPhone. In order to address this concern, which she believed would potentially make her videos less likely to be consumed, as they would be of lower quality, P3 altered her creative routine to draw upon other video-recording resources, stitching together her professional camera and TikTok's infrastructure to seamlessly reengage in her creative routines. This allowed her to upload higher quality videos that she believed her followers were more likely to engage with. As P3 makes videos about coping with ADHD for herself and for people like her, she felt that having polished videos may make them seem more accessible, and therefore helped her help others.\par

Many of our participants discussed their own neurodivergence, citing autism, ADHD, and other attention-related challenges that made creating on TikTok inaccessible. P10, who has both ADHD and autism, made this connection explicit when asked what some of the challenges they were having with TikTok.  
\begin{displayquote} "Allowing people to pause the video while they're doing editing so that they don't have the cognitive load of having to hear themselves over and over again, like that alone. Why can I pause a video I'm watching but I can't pause the video I'm editing?" \end{displayquote}
Sensory processing issues like what P10 describes as 'cognitive load', are commonplace for people with autism or ADHD (P10 reports having both), and can often compound or impact multiple senses at once \cite{baum2015behavioral,sjowall2013multiple}. The visual stimuli, as well as the auditory feedback, while attempting to edit a looping video on TikTok rendered TikTok incredibly challenging for P10 to use. The over-stimulation of hearing one's voice, as well as attempting to engage in a largely visual and physical task while editing, led to frustration \cite{seymour2019frustration} and self-doubt \cite{sjowall2013multiple}. (P2, who also has autism, also made a similar point, and P3's point above echoes the same sentiment). P10 continues, reflecting on how their struggles at editing has impacted their creative routine: 
\begin{displayquote} "And so, TikTok, without that immediate feedback, I'm just hearing myself and [...] judging myself pretty harshly. [M]y therapists actually advise me to do this, just like, make the first video and throw it away. And then just go for the second one [...] and be okay. And as somebody who never got to spend time being creative, [...] I've lost a lot of practice at failure as a creative person. And so my creative process still feels very clunky and unnatural."\end{displayquote} 
The work and cognitive load involved for P10 to enact their creative routines directly in TikTok's interface makes creation challenging and inaccessible. \par

For P10, despite their complaints about how use of Adobe Premiere presented a barrier to both success and access, this meant contemplating altering their creative routine to learn and integrate Adobe Premiere into their creative routine. They explain: 
\begin{displayquote}
"I would love to spend the time to learn premiere because I really struggle with switching back and forth. Like anything with any kind of complex editing."\end{displayquote}
By looking to make the video editing easier in order to lessen the cognitive load of editing multiple clips (the switching back and forth that P10 describes), P10 is attempting to draw on multiple infrastructures in order to continue to engage in their creative routines. By looking outside of TikTok's infrastructure for ways to better enact their creative routines, P10 is starting to engage in bricolage tactics to repair a gap in TikTok's infrastructure that has made the platform challenging, and sometimes inaccessible for them to  use. These practices, as their comment around "just hearing" themselves so clearly illustrates, will also allow P10 to engage in their creative routines while drawing on external infrastructures with a lighter cognitive load. \par

Our participants drew on external software (P4, P10, P13) and hardware (P3, P15) in order to address gaps they saw in TikTok's infrastructure that made engaging in their creative routines challenging or, at times, inaccessible to them. P4 altered her creative routine to stitch together an external editing software and a file sharing network in order to be able to enact her creative routines on TikTok. Despite this critical infrastructuring to repair the inaccessible parts of TikTok's infrastructure, P4's solution rendered other aspects of TikTok inaccessible, as she was not able to use the stitch feature on account of she processed things in a non-linear fashion and still enact her creative routine. P10, in contrast, found that editing in TikTok's interface was challenging to them as it overstimulated their senses and made enacting their creative routines challenging. Their solution of incorporating the act of 'throwing away' recordings of themselves and 'being okay with that' making their creative routine feel 'clunky and unnatural,' which subsequently led to P10 looking to alter their creative routine further to learn an external video editing software. These critical infrastructuring practices allowed our participants to draw upon multiple infrastructures in order to continue to enact their creative routines, while also presenting new and unexpected challenges to access. 

\subsection{"Hey Can You Add Captions?": Expanding TikTok's Capabilities as Infrastructuring Care}
As our participants spent more time on TikTok and settled into routines around making videos, their concerns around the platform started to shift. During our conversations around their creative outputs and the emergent challenges of making and sharing videos on TikTok, our participants shared stories of how they'd either established, or altered, their creative routines around making videos on TikTok in order to ensure that the videos they made were accessible to people who watched them. This, at times, reflected their personal preference for how they themselves wanted to consume TikTok videos. Unlike the previous section, where critical infrastructuring to address infrastructural gaps was largely focused on how challenging TikTok's infrastructure was for neurodivergent people to use, here our participants collectively observed a gap in TikTok's lack of infrastructural support for closed-captioning on videos, which could potentially render the outcome of their creative routines inaccessible to  audiences. More than half of our participants (P2, P3, P4, P8, P10, P11, P12, P13) mentioned their neurodivergency as a motivating factor in why they chose to expand TikTok's capabilities and caption their videos. Previous work has shown how subtitles and closed-captioning can help people with ADHD and other attention-related or sensory processing challenges with information retention and learning \cite{lewis2012multimedia}. \par

For some of our participants, the decision to add captions to their videos was a consideration not just for the audience, but rather a reflection of their  viewing preferences that they had integrated into their creative routines. P14 explains: 
\begin{displayquote}“[F]or me, I always think whenever I consume anything, I would like to see [captions]. I also noticed that sometimes I wasn't able to engage with people's videos like I wanted to. […] So I figured, like, accessibility is key. \end{displayquote}
P14's decision to add captions to her videos reflected the value she herself placed around being able to consume videos on TikTok in a way that allowed her to follow what was going on. While P14 did not disclose that she was neurodivergent over the course of our conversation, she describes her inability to engage with TikToks without captions in ways that closely align with the needs of neurodivergent people \cite{lewis2012multimedia}. Moreover, P14 drew on TikTok's infrastructure to add captions directly to her videos,  noting that adding them allowed her  the benefit of being able to clarify what she said in videos by adding nuance in the captions.\par

While P14 was able to seamlessly integrate TikTok's existing infrastructure into her creative routine not only around making videos, but also captioning them, this was not the case for our neurodiverse participants. Many of our participants said the reason they wanted to see captions on videos, and why they took great pains to ensure that the videos they made had captions despite the challenges of adding them, was because they themselves were neurodivergent and needed video captions to enjoy TikTok \cite{lewis2012multimedia}. For example, P12 explains how adding captions to her videos is challenging on account of her ADHD, but she adds them to avoid outing herself as neurodivergent:
\begin{displayquote}"I have ADHD, but that's not a part of my identity. It's not a big deal. [...] [S]eeing someone or [a] mutual [say] like, Oh, they can't add captions. And so they have to out themselves as being neurodivergent. So that you forgive them. One of my biggest, I guess, triggers is when I'm doing a caption and I can't. [...] I start panicking and I throw my phone, and like, one day, I'm gonna break it."
\end{displayquote}

Adding captions disrupted P12's creative routines, as she was not able to insert them easily, much in the way P3 described editing video clips together on TikTok's native editing suite in the previous section. The work involved in adding these captions further frustrated P12. Yet, P12's commitment to adding captions, despite the challenge adding them presented to her because of her ADHD, was not unique in our conversations. Over half of our participants (P2, P3, P4, P8, P10, P11, P12, P13, P15) discussed how they have integrated adding captions into their videos as a part of their creative routines despite the challenges they encountered doing so. This, according to many of our participants, was out of consideration for their audience beyond those whose neurotypes and ability status align with their own. As P15 explains:
\begin{displayquote}"[T]here are like so many people that are like [d/D]eaf or hard of hearing and it is so frustrating. [...] Whatever I record, I will put in captions, but it's so time consuming, it takes so much [effort] and I'm gonna do it anyway because it is a necessity. Yeah, it would be so nice if TikTok automatically did auto captioning, because people need that, like, it should be definitely more deaf accessible, and like just disability accessible." \end{displayquote}
Both P12 and P15's experiences reflect how hard adding captioning is using TikTok's editing suite directly. They also reflect a commitment to a community norm they both observed and benefited from on TikTok, as the push to caption videos for purposes of ease of consumption and access to everyone emerged organically on TikTok. P15's comment goes a step further and specifically calls out the infrastructural gap she perceived on TikTok, where, at the time, there was no automatic captioning feature included in TikTok's infrastructure. \par

Some participants also sought to raise awareness around how difficult it was to work within TikTok's infrastructure to add these captions by inviting viewers of their videos to try their hand at it. P13 created a video discussing captioning and the challenges of captioning videos using TikTok's available infrastructure. They explain,
\begin{displayquote}"The premise of the video was anybody who has the time, anybody who wants to try it: Take this video, duet it or download it or whatever, however you want to do it, and caption it yourself. Come back and tell me what the experience was like. I wanted more people to understand what it is like, how much work captioning is because people who don't caption anything, because they don't make content [they have] have no fucking clue."\end{displayquote}
P13’s challenge to their followers was meant to demonstrate how TikTok's existing infrastructure presents monumental challenges for neurodiverse video makers, as well as those who experienced similar struggles or wanted to support those who did, to create accessible videos or to easily integrate the practice of adding them into their everyday creative routines on TikTok. P13's commentary also carries with it the implication that integrating captioning into one's creative routine creates additional, challenging, labor. P13 13 goes on to add:
\begin{displayquote}"Because [...] manually captioning is an accessibility problem for a lot of people. Like, there are a lot of content creators who have like muscle disorders that make it difficult for them to type. And captioning is like prohibitively difficult for them. And [...] when we put the burden of making content accessible on those creators, [...] we're making the possibility of being a content creator less accessible."\end{displayquote}
Not being able to meet the community norm around integrating captioning into one's creative routine of making videos on TikTok can make it harder for video makers to have their work engaged with on the platform. P13's video challenge to others was to raise awareness of how much harder it can be for people. They draw on people with different physical abilities, but this also extends to people, like P13, who are neurodiverse as well. \par

Yet, despite how captioning could be considered inaccessible, our participant's efforts to add captions anyway demonstrate a clear care for other users of the platform. It further shows how infrastructures can be generative through a community's collective experiences \cite{star1996steps}. Put another way, the ability of everyone to use and create with TikTok's infrastructure is an evolving conversation between the audience and the creator. Our participants are engaging in critical infrastructuring practices to ensure their videos meet their own needs and in doing so, are also attending to the needs of others.  \par

Our participants, in order to get around the work and challenge that captioning their videos within TikTok's existing infrastructure provided, drew upon other applications to fill this infrastructural gap. The applications CapCut (P5, P15), Instagram Threads (P2, P5, P6, P7, P15) and other services such as those integrated with Adobe Premiere (P4, P13) were all discussed by our participants. Many participants told us that they saw captioning as a necessity for accessibility of their videos by the larger TikTok community, as well as to meet their own needs as neurodiverse users to TikTok. Some cited their own desire to view captioned content (P4, P6, P7, P13, P14, P15), others cited accessibility concerns for the broader TikTok community (P10, P11, P13, P15), personal connections to d/Deaf and hard of hearing family members (P2, P3), or because viewers specifically requested that they caption their videos (P5). While their reasons to add captions to their videos are multiple and complex, the majority of our participants started adding captions to their videos directly in TikTok’s in-app editing suite. The expansion outward to augment their existing creative routines came from a want for better ease of use, or because the process of captioning within TikTok’s editing suite was challenging for them, despite their dedication to engaging with the community norm around ease of access and consumption of their video content. \par

While our participants relied heavily on external infrastructure like CapCut or Threads, TikTok’s internal infrastructure presented challenges to universal use of these external applications, similar to the editing challenges discussed in the previous section. P2 explains:
\begin{displayquote}“When I am responding to something---like if you stitch something or duet something---you can't use the captioning app. Because then it will take away from the connection the video has to [another] user, so I have to type it in by hand. [...] So it takes me like an hour to do that.”\end{displayquote}
Duets, replies\footnote{where the creator makes a video to respond to a comment on one of their replies}, and stitched videos cannot be recorded or edited outside of TikTok’s native editing suite, as P4 also noted in the previous section. This is, in part, because TikTok’s internal archival structure requires that stitched and duetted videos directly link back to the original post by the initial video’s creator. The stitch and duet features are commonly used features on TikTok, and are broadly used for participatory challenges, e.g. the “tell me you’re something without telling me you’re something” challenges that users will stitch with the original video. The expanded infrastructure our participants built outside of TikTok that allowed their videos to be more accessible to their viewers and more accessible to their own creative routines, cannot be used with videos that are in conversation with other users or video content, making these creative forms more inaccessible to our creators.\par

\subsection{The Pitfalls of Top-down Infrastructural Solutions: Loss of Agency and Control with TikTok’s New Auto-caption Feature} These interviews took place between February and March of 2021. In April 2021, TikTok rolled out a new auto-captioning feature. In the process of sending copies of their interview transcripts to our participants, we asked a follow up question to participants regarding their use of, and thoughts about, the new Auto-Caption feature. Out of our fifteen participants, five (P2, P5, P7, P8, and P10) responded to our follow-up request. All that responded, with the exception of P8, reported using the Auto-Caption feature, which had not yet been released to them. Despite this, P8 reports: 

\begin{displayquote}“I think [auto captions] could be better as I have seen a lot of people just use it without editing the caption so occasionally words in the captions are wrong.” \end{displayquote}

In addition to also noting that the auto-caption feature supplied inaccurate transcription, P10 also made a note about how TikTok's auto-caption feature “censors words without explanation.” Going back in and altering what was auto-transcribed was seen as creating more work for our participants as creatives, and, for some, not worth the effort.\par

Another frustration our participants voiced was with the way that TikTok's auto-caption feature took a way a lot of control around how their captioning efforts looked on their videos. P2 explains:

\begin{displayquote}“I thought the feature would be more unique but it's almost identical to Threads in terms of use. The only thing I don't like about the feature is that I can't choose the placement of the captions like I could with Threads. Choosing placement, being able to change the font or even color would help the captions be easier to read.”\end{displayquote}

While our participants were using the new feature, it was clear that they still wished for more control to more purposefully design their video content. P2’s comment about placement of captions, including font, text color, and placement of the captions on the video screen, shows how much control our participants had through their stitched infrastructures. Despite auto-captioning now existing on TikTok, our participants feel TikTok still did not, fully, make these videos easier for our participants and their audiences to consume, or for our participants to create and edit in TikTok’s in-app editing suite. While TikTok did add a feature that added accessibility to videos produced and edited within their application, it did not fully meet the community’s needs or requirements around such an infrastructural tool.

Our participants also discussed a loss of control and nuance in what they were providing in their captioning when discussing auto-captioning. We see this in the censorship P10 described, and the inaccuracies that P8 talks about, as well as in P2's frustrations with with her loss of control over \textit{how} captions were integrated into the videos she produced. Another loss of control our participants described spoke more to the care for the community that adding captions came to embody for our participants. P7 explains what he saw as lacking in the new auto-caption feature,
\begin{displayquote}“[Auto-captioning] does not do visual description for screen readers, which I put in comments (along with tone indicators).”\end{displayquote} 
The frustration our participants felt with this new feature comes, in part, because of how TikTok's auto-captioning feature was designed with a different intention than the reason why they chose to add captions to their videos. While their captioning efforts with all their requisite stitching of disparate infrastructures together were done on behalf of the community to generate a bottom-up community norm around creating a more accessible TikTok through things like visual descriptions for screen readers or tone indicators,\footnote{Tone indicators are short-handed statements at the end of a sentence (e.g. "-s" for sarcastic or "-gen" for genuine). They are generally used in text-based communication where tone/nuance can be lost. A full list of tone indicators can be found here: https://toneindicators.carrd.co/\#masterlist} TikTok's auto-caption feature lacked all of these nuances. The one-size-fits-all solution TikTok provided meant that many of the additional considerations for the community our participants made in adding captions to their videos were lost, and still had to be repeated in addition to correcting the captions and ensuring they would not be censored. For all of our participants, their creative routines had to be altered again in order to integrate this new feature. This created new challenges and frustrations that, despite the convenience of having auto-captioning built into TikTok, created additional steps for our participants to continue to engage in the critical infrastructuring practices that produced bottom-up, community-driven norms around captioning that TikTok's top-down solution failed to provide.

\section{Discussion}
In this section we discuss two major observations from our findings. The first is focused around the potential byproducts of critical infrastructuring work. While our participants initially engaged in infrastructuring to maintain and sustain their creative work, we see that the byproducts of critical infrastructuring are explicitly tied to care and repair work. To illustrate this, we develop the concept of incidental care work, where we consider how critical infrastructuring is not merely a reflexive practice that leads to broader critique of social systems and arrangements, but rather a practice that can have unintended pro-social effects through its enactment. In the next section, we consider accessibility in the context of critical infrastructuring, with a particular focus on context and how accessibility is not a static construct through a concept we dub contextual accessibility. Here, we highlight how accessibility is a situational, co-constructed concept defined by the actors who are working to make things accessible at any given time. 

\subsection{Incidental Care Work: Critical Infrastructuring and it's Byproducts}
One of the emergent themes in our findings is a discussion around care, specifically the care that goes into the work of creativity. Creativity is one of those mushy concepts that is always ill-defined in the literature, but as Do and Gross \cite{dogross2007creativity} imply, one of the central tenets around creativity is that \textit{something} is produced as a byproducts of that labor. The videos our participants produced are prime examples of creative labor. Returning to the concept of the produser \cite{bruns2006towards}, creative labor in online spaces requires that the user must co-produce the platform and by extension the \textit{platform experience} upon which they labor. While an individual on a platform that plays host to millions of users may not be able to directly shift or alter the platform experience overall, our findings show how, on TikTok, there is a firmly held belief that the creator is, at least in part, responsible for creating the platform experience for others as it relates to their creative work. A produser, then, becomes ruler of the fiefdom of their content; and it is their care and responsibility that structures, regulates, and creates the infrastructures of accessibility within their creative work.\par

In assuming this responsibility, many produsers shift or alter their routines around their everyday creative practices. These routines are also products of creative labor. Many of the experiences described in our findings can, on the one hand, be seen as our participants engaging in critical infrastructuring to make their creative labor easier to produce and perhaps more enticing to viewers, garnering them positive feedback in the form of views, likes, and comments on TikTok. We see examples of this in how participants were willing to caption their videos, but very quickly looked externally to TikTok to find less labor-intensive solutions for them to incorporate into their routine creative practices. On the other hand, these practices are rooted in care for their viewers and themselves, and the byproduct of that care is the critical infrastructuring work that allows for accessibility. \par 

Critical infrastructuring, as described by Britton and colleagues, is at once a reflexive practice and a practice that critiques broader social systems and arrangements \cite{Britton2019}. In this case, what started as a reflexive practice to better scaffold and support the creative labor of our participants on TikTok incidentally turned into broader care for the larger TikTok community. As the infrastructural logics of TikTok did not support many of our participants' creative routines, and at times rendered them inaccessible, participants engaged in a process known as repair work in order to make accessible and enact their creative routines. Repair is the "subtle acts of care by which order and meaning in complex sociotechnical systems are maintained and transformed, human value is preserved and extended, and the complicated work of fitting to the varied circumstances of organizations, systems, and lives is accomplished" \cite[~p.222]{jackson2014RethinkingRepair}. Care here is of particular note, because a person must \textit{care} if they wish to engage the critical infrastructuring needed in order to efficiently innovate an existing infrastructure \cite{jackson2014RethinkingRepair}. The acts of care needed to stitch together disparate infrastructures in order to engage in routine practices of creative labor highlight how sometimes, the byproducts of these actions can have a net benefit to a platform community. \par

While the produser on TikTok is only responsible for their own experience on the platform, the byproduct of their bricolage practices to address perceived infrastructural gaps is a broader, platform-wide norms around how to best create easy to consume and understand video content. We see this in how our participants routinely worked to care for their audiences through their video content, and TikTok's platform interface itself accessible to both them and anyone who wished to interact with them or their content. They altered their routines around their creative labor to produce \textit{accessible} video content in a routine exercise of care. While our participants came from many different backgrounds, their needs as neurodiverse people, particularly people with ADHD, shaped this care despite the challenges it presented to our participants on account of their neurodiversity. In the subsequent two sections we outline these practices of critical infrastructuring as they relate to their results and contribute to the larger conversation in showing how, at times, this critical infrastructuring work can emerge in subtle ways as the byproducts of seemingly self-centered actions.\par

\subsubsection{Infrastructure for the Self: Reflexive Practices} The principles of universal design suggest that products should be able to be used by everyone without the need for accommodation or specialized design \cite{story2001principles}. Yet, as our participant's stories show, platforms like TikTok are designed to cater to an able-bodied and neurotypical userbase. The top-down infrastructure of TikTok does not provide the produsers, upon whose creative labor the platform depends, with the necessary tools and infrastructures to allow for the routine enactment of creative labor. By looking elsewhere to find solutions to these infrastructural shortcomings, our participants reflexively navigate the demands of being a content creator in a space like TikTok when that space is inaccessible in various ways on account of their neurodiversity. We see this in how our participants looked to external infrastructures to make accessible editing and recording videos, in their creation and use of captions, as well as how our participants developed new routines around creating accessible and stable spaces for community through their added consideration towards the larger TikTok community through the addition of tone indicators, or concern over the placement of captions within the body of their videos. However, these actions were not without their potential downsides. Oftentimes the critical infrastructuring involved to make certain aspects of TikTok's existing infrastructure accessible so our participants could enact their creative routines as neurodiverse creators rendered other aspects of TikTok inaccessible, or presented new challenges. As our participants navigated these challenges, they had to evaluate and make decisions around what they were willing to sacrifice in order to have a positive experience on the platform. These cost-benefit analyses demonstrate the reflexivity needed in order to engage in critical infrastructuring on a platform like TikTok.

\subsubsection{Infrastructure for Others: Critique in Service of the Broader Community:} A byproduct of the reflexive practices and what is given up or gained by creators on TikTok as they engage in critical infrastructuring is an outward critique of the existing platform infrastructure and broader TikTok community. This is best illustrated through how our participants looked beyond TikTok's infrastructure to repair and make accessible the outputs of their creative labor by looking for external infrastructures that would allow them to add closed-captioning to their videos. Here, we can see how the efforts to caption demonstrate both an adherence to an emergent normative practice among members of the TikTok community broadly (to make content accessible with captions), and a care for the viewing experiences of people who may watch their videos. This  demonstrates how critical infrastructuring can lead to incidental care work. When one thinks about accessibility frameworks like independence, such as providing the tools and structures to allow people to dictate their own experiences \cite{InterdependenceAsstTech}, one can further see how context in this case really matters to conversations about accessibility on TikTok. Who is creating the accessibility? What is being made accessible? And through what practices is this being done? In the next section, we build and elaborate on these questions in a discussion around the accessibility that these critical infrastructuring practices provide to both individuals and the general TikTok community.

\subsection{Accessibility in Context: Contextual Accessibility}
One of the key findings from our results are the ways in which creators routinely engaging in creative labor on TikTok occupy a space where they are both improving upon, but also creating within TikTok’s existing, incomplete, infrastructure. The critical infrastructuring practices creators are engaging in are derived from the routines and practices that both shape and are shaped by their navigation and bricolage repair of TikTok's existing infrastructure. Outside of the incidental engagement in care work as discussed in the previous section, another byproduct of these critical infrastructuring practices is the access that emerges through the accidental care work to create accessibility -- both for creators and for others -- within the context of that infrastructuring work. Our findings show how, while maybe not the stated goal of critical infrastructuring actions, access to community spaces and the creation of accessible content and bricolage development of accessible infrastructures, are byproduct of individual actions taken in various contexts on TikTok. Collectively, these findings show how context of what accessibility means, as well as the means by which that accessibility is created, are fundamentally important to conversations about accessibility broadly. 

Accessibility in the literature is commonly defined as a form of labor to build access that centers disabled people \cite{InterdependenceAsstTech, hofmann2020living, bennett2020care, hamraie2019crip} through a community oriented approach \cite{das2019doesntwinyoufriends,2016ComicTranscriptionsFiesler}. These community-driven and bottom-up accessibility efforts are seen through the use of participatory methods \cite{Abouzhara2013webacc,kuksenok2013,Garcia2009} or when users revise existing systems to address accessibility issues on their own terms \cite{Low2019twitter}. Moreover, the work of infrastructuring for accessibility is often invisible and engaged in by people with various disabilities \cite{Branham2015Invisiblework}. In this sense, accessibility is also an evolving construct that is continuously shaped \textit{in situ}. Put another way, the context of the work of accessibility matters; as a labor practice, accessibility is not stable.

Through our analysis, we have observed how accessibility is not always the stated goal of individual actions, however it can be the byproduct of critical infrastructuring efforts to address observed gaps in existing platform infrastructure. Oftentimes, however, accessibility as a byproduct of critical infrastructuring actions is an imperfect solution that sometimes complicates or introduces new accessibility challenges for people. Moreover, we find that people on TikTok make normative assumptions about the platform and the TikTok community beyond their network of viewers and followers. The collective decision by many of our participants to do the work of accessibility to alter their creative routines to add closed-captioning to their videos (despite how inaccessible doing this may be to them), is derived from an assumption that most people on TikTok would like there to be captions on videos. While myriad community norms are communicated to creatives on TikTok through their routine engagement with the platform, the norms around "accessible content" were shown to drive critical infrastructuring practices that draw on infrastructures both internal to and external to TikTok. \par

We find that creators attempt to collect, coalesce, and condense their own definitions, as well as individual audience members’ and followers’ definitions of what makes videos \textit{accessible}, according to their individual needs, into tangible \textit{acts} of accessibility (e.g. captioning). We further find that TikTok creators draw upon their own wishes for what they’d like to see in content and reflect those back to the community. We define this sort of accessibility on TikTok as \textit{the cyclical practice of making videos interpretable to users, regardless of ability or environment, that is constantly updated as new audience members expose makers to their unique needs.} The reflexive ways creators are engaging in individual acts of accessibility and drawing on community norms to define accessibility in context are contextual accessibility. \par

As a conceptual construct in HCI, context has transformed from the positivist perspective of the setting where action unfolds, a stable entity disconnected from the activities taking place therein \cite{dourish2004we}. The formative work of Lucy Suchman \cite{suchman1987plans}, however, integrates the activities of humans into the concept of context, illustrating how people’s activities are neither stable nor predetermined, and how context is dynamic and always changing. Context cannot be determined \textit{a priori}, rather it is an emergent property of interaction \cite{dourish2004we}. In other words, context is a continuous production, determined by the people who are present in a space and in how they collaboratively generate the rules and norms for a given interaction. In our work, we see how accessibility is also a contextual construct, actively produced and determined by members of a community. Importantly, our participants—creative content creators—possessed a kind of infrastructural competence \cite{sawyer2019infrastructural}, in that they worked to bring together various tools and experiences in useful alignment to address issues of accessibility. However, this competence was not simply an individuated activity, but rather a collective one that relied on their ability to align their own histories and experiences with accessibility as neurodivergent creators with the normative expectations of accessibility within their online community space. \par

Our results show that for creators it is critical that TikTok videos be accessible to viewers and potential viewers. Without this accessibility, there is a perceived lack of solidarity with the larger platform community. This is particularly true for creators who find consuming content without accessibility considerations challenging themselves as neurodivergent creators. Our results show how expanding audiences allowed our participants to critically assess the videos they shared on TikTok. For others there was a desire to create, in their own videos, content that they themselves would find accessible. P4 and P14's examples above is illustrative of this, where their own want to see captions or subtitles on videos was driven by their own media consumption routines. The contextual accessibility developed by our participants comes from the collective, but it is also informed by personal desires and embodies incidental care for the community.\par 

\section{Limitations and Personal Reflections}
This research was not designed as an accessibility or disability research study. We initially set out to speak to people who made videos and shared them on TikTok, with a hope of learning more about their creative practices and routines. As such, our inductive findings led to questions during the review process regarding whose voices were and were not included in our research. Through our inductive approach, we learned that the majority of our participants were neurodiverse people, people with closely-aligned experiences, and people whose actions tacitly supported the neurodiverse community. They engaged in creative labor and emergent practices to address issues with TikToks infrastructure in supporting people like themselves in creating as well as in engaging with creative content. To address this limitation, we are careful to limit the scope of our project to focus largely on neurodiverse people, particularly those with ADHD, and those with similar experiences, as the everyday life experiences of people with ADHD are often left out of conversations like this regarding ease of use and access to technology \cite{SpielADHDTech2022}. We note and regret that we did not speak to anyone from the D/deaf community. Because of this, we frame our work so as not to speak for them and their accessibility concerns, despite the overlap and obvious benefit that community has from the critical infrastructuring practices our participants discussed. 

We adopted the frame of critical infrastructuring as this form of infrasructuring was intentionally conceived as a way to draw explicit attention to the bottom-up practices of people that revise or push-back against top-down normative efforts and logics \cite{Britton2019}. This allows us to speak more broadly to the concerns of the TikTok community regarding norms around access and ease of use, noting how these infrastructuring practices produced structures and support for people whose needs may be different than what the platform itself provides, even after community-driven pushes for improved features \cite{InterdependenceAsstTech,Odell2020}. The infrastructuring practices described in this paper around access and ease of use described in this paper are beneficial to neurodiverse people, a community whose everyday, routine encounters with technology in creative contexts around visual media is understudied or largely presented in ways that frame such divergences as a deficit\cite{SpielADHDTech2022}. The findings in this paper point to a need for further research into the d/Deaf and Hard of Hearing community's organizational efforts and infrastructuring practices on TikTok to enforce bottom-up development of community norms for support, as well as a need to further examine the creative processes and everyday routines around creativity, as well as concerns around access and ease of use, that neurodivergent people have on visual-media sharing applications like TikTok. \par

While writing this paper, the first author was diagnosed with ADHD. Research into creating infrastructures for access and ease of use -- accessibility -- that can benefit more than just one community are often singled out in HCI as only being "for" one particular community. This is troubling to us as researchers whose inductive analysis produced such findings. Community-led efforts around accessibility informed by friends, by family, and by one's own personal preferences in the consumption of visual media are vital, as they all contribute to a structure and system that allows for both independence and interdependence \cite{InterdependenceAsstTech}. 

By doing critical infrastructuring through routines of creative making, the broader TikTok community is making the platform accessible to everyone in both a social and technical sense, and these efforts are not just made for those who may be the most obviously in need of these structures of access. This is a different take on notions of universal design \cite{iwarsson2003accessibility}, designing to make the world usable and useful by as many people as possible, whereby access is a byproduct of routine practice. Moreover, the access observed in these findings is created and made by and through the care people who engage in creative labor of producing user-generated content put into ensuring that their platforms are both technically and socially accessible to anyone who may chance upon the end result of that labor.  \par

This paper has several other limitations. Firstly, the first author’s recruitment video was shared via TikTok's duet feature by several TikTok users within tightly-knit communities (particularly around Math TikTok), meaning that while the video did diffuse across multiple networks, these networks were small and potentially limited. By recruiting participants from other social media platforms, as well as through several Discord Servers, an effort was made to include individuals from other networks. A second limitation emerged around a small sample size, as we wanted to ensure we had a diverse sample. We did not want to represent a self-identified white, cisgender, and female sample exclusively, so we selected toward a more diverse sample, which meant that the overall number of participants sampled was small. \par

The authors of this paper come from myriad backgrounds. The first author, who conducted the interviews and conducted the bulk of the analysis, is a white, cisgender queer woman. In order to mitigate any unconscious bias that may have occurred over the course of the analysis on account of her whiteness, the first author consulted regularly with the second and third author. The second author is a cisgender woman child of immigrants from a multi-ethnic background. She primarily contributed to the literature review and helped to develop the arguments made in the discussion. The third author is a cisgender Middle Eastern man, who is a member of a minority religious group. He served as the anchor author, guiding and shepherding the work from its inception.\par

\section{Conclusions and Future Work}
In this piece we examined how TikTok creatives are engaging in acts of critical infrastructuring that are at once reflexive while also serving as a critique of the broader TikTok community and infrastructure. An interesting and observable byproduct of these critical infrastructuring practices lies in how they create and maintain structures of accessibility that are both emergent and reflexive. We discussed how at the heart of many of these self-serving infrastructural practices oftentimes resulted in a more altruistic result in the form of care work. We further show how the context of these critical infrastructuring practice---something that grew out of both creatives themselves as well as the collective userbase of TikTok---is important to how they are assessed, particularly around discussions of access and ease of use. \par

Infrastructuring is a community practice. Not only were these creatives were engaging in these critical infrastructuring practices on their own; but rather, people from various identities and abilities were providing input and collaborating on the infrastructuring practices described in this paper. As researchers, we did not purposefully include people from some of these identities because that was not the original intention of the study. Reflecting upon this, it is clear how this is both a shortcoming and a potential strength of this research. On the one hand, this paper allows a window into the everyday routines of TikTok creatives, and shows how though engaging in critical infrastructuring practice people who maybe do not fit the bill of general accessibility activism and advocacy are, perhaps accidentally, engaging in care work for the community that improves access broadly. On the other hand, it is clear that there are voices lacking in this piece, and that these present avenues for future work. Our work contributes a deeper understanding of the creative practices engaged in by creatives with ADHD and those who whose actions support neurodiverse people on TikTok in making a digital platform accessible on their own terms. We wish to continue the call to action initiated by Spiel and colleagues \cite{SpielADHDTech2022}, encouraging others in our community to explore the everyday practices of people with ADHD. We do wish to be clear in saying that those with ADHD are not more marginalized than other groups. Rather, they are an understudied population whose voices and experiences can lead to a foundational understanding that can yield more inclusive future technologies. By understanding the bottom-up practices of people with ADHD, we can begin to develop a broader understanding of what works for people on the ground as opposed to engaging in top-down design practices. \par

In considering the individual acts of community enrichment through both their own content and the broader TikTok community, a potential avenue for future work lies in examining how platform actions can frustrate and complicate the moderation aspects of individual users who may be moderating a community of one - or a community around their own creative content. An exploration of specific community needs and infrastructuring practices on TikTok, such as the neurodivergent or d/Deaf and Hard of Hearing communities, would help to better understand how creatives on TikTok are working to improve and build accessibility for TikTok through critical infrastructuring.

\begin{acks} We'd like to thank our participants, without their insight none of this work would be possible. We'd also like to thank the members of the al-adala lab at CU Boulder for their thoughtful feedback, as well as our anonymous reviewers.
\end{acks}

\bibliographystyle{ACM-Reference-Format}
\bibliography{CSCW2023_Final}

\end{document}